\documentclass[12pt]{article}
\usepackage[utf8]{inputenc}
\usepackage{natbib}

\usepackage{balance} 

\usepackage{algorithm}
\usepackage{algorithmic}
\usepackage{nicefrac}
\usepackage{epigraph}

\usepackage{amsmath}
\usepackage{amsthm}
\usepackage{amssymb}
\usepackage{bbm}
\usepackage{bm}
\usepackage{mathrsfs}
\usepackage{amsfonts}

\usepackage{blkarray, bigstrut}
\usepackage{enumitem}
\usepackage{adjustbox}
\usepackage{subfigure}

\title{Pandering in a (Flexible) Representative Democracy}

\author{
  Xiaolin Sun\\
  \texttt{xsun12@tulane.edu}
  \and
  Jacob Masur\\
  \texttt{jmasur@tulane.edu}
  \and
  Ben Abramowitz\\
  \texttt{babramow@tulane.edu}
  \and
  Nicholas Mattei\\
  \texttt{nsmattei@tulane.edu}
\and
  Zizhan Zheng\\
  \texttt{zzheng3@tulane.edu}
}

\newcommand{\BibTeX}{\rm B\kern-.05em{\sc i\kern-.025em b}\kern-.08em\TeX}

\newtheorem{theorem}{Theorem}

\newtheorem{problem}{Problem}

\DeclareMathOperator*{\argmax}{arg\,max}

\begin{document}

\maketitle

\begin{abstract}
In representative democracies, the election of new representatives in regular election cycles is meant to prevent misbehavior by elected officials and keep them accountable to the ``will of the people."
This ideal is undermined when candidates are dishonest when campaigning for election over these rounds. Much of the work on Computational Social Choice to date has investigated strategic actions in only a single round.
We introduce a novel formal model of \emph{pandering}, or strategic preference reporting by candidates, and examine the resilience of two voting systems, Representative Democracy (RD) and Flexible Representative Democracy (FRD), to pandering within a single round and across multiple rounds.
For each voting system, our analysis centers on the types of strategies candidates employ and how voters update their views of candidates based on how the candidates have pandered in the past.
We provide theoretical results on the complexity of pandering in our setting for a single cycle, formulate our problem for multiple cycles as a Markov Decision Process, and use reinforcement learning to study the effects of pandering by both single candidates and groups of candidates over many rounds.
\end{abstract}





\section{Introduction}

Modern representative democracies use regular elections to ensure that officials uphold the ``will of the people." Periodic elections are meant to prevent corrupt or ineffective officials from maintaining power and to keep them honest. However, current electoral systems are arguably insufficient for this task because voters only have a say during the election (aside from the potential recalls), which typically occur during regular cycles or \emph{rounds}. 
In recent years, ideas of \emph{delegative} voting, including liquid democracy \cite{blum2016liquid} and flexible/weighted representative democracy \cite{abramowitz2019flexible,weighted} have been advanced in the computer science and social choice literature. In these delegative voting systems the voters collectively weight their representatives, possibly updating their weights during a round, i.e., between elections. These systems typically fall under the heading of \emph{interactive democracy} \cite{brill2018interactive}. 

It has been proposed that delegative voting schemes can interpolate between direct and representative democracy and would be better at keeping representatives accountable~\cite{abramowitz2019flexible,ford2002delegative}. However, the idea that delegative voting will be better at keeping representatives accountable, or that it will be better at expressing the ``will of the people", is largely untested aside from some nascent applications of Liquid Democracy (with transitive delegations)~\cite{paulin2020overview,civicracy}. Little is known about how such systems perform in the presence of agents who are strategic, selfish, and even malicious~\cite{zhang2021power,bloembergen2019rational}. Answering questions surrounding responsiveness to voter preferences and robustness to bad actors is critical for selecting and comparing various democratic systems.

One of the primary features of representative systems is that candidates campaign for votes, making promises and promoting platforms of action on decisions to be considered in the future. This process of campaigning is important to inform voters of their options and seek support for these promised positions. Unfortunately, politicians lie, especially when trying to get elected or maintain power. This \emph{pandering} is a form of attack on representative democratic systems, and we introduce the first formal model of pandering to the literature on Computational Social Choice (COMSOC), which has previously considered other forms of election attack including manipulation, bribery, and control \cite{brandt2016handbook}.

Pandering during campaigns is not a new problem, it is a global phenomenon. US citizens consistently rank Congresspeople as occupying the least trustworthy profession \cite{mccarthy_2021} 
and over half of Americans are unsatisfied with representative democracy as it stands \cite{wike_schumacher_2021}. A study involving Spanish mayors \cite{janezic2020eliciting} demonstrated that lying may increase a politician's chances of being reelected. Studies have shown that voters are often aware of pandering and become suspicious of politicians that are perceived as panderers \cite{pandering_politicians_suspicious, mcmurray2017polarization}. Australian voters demonstrably decrease support for politicians upon the revelation of their lies \cite{aird2018truth}.

Our core concern is whether delegative voting systems are more or less vulnerable to dishonest candidates. To study these questions we model two types of democratic voting system; classic Representative Democracy (RD) and Flexible Representative Democracy (FRD) \cite{abramowitz2019flexible}.
For each system, each election cycle consists of (1) voters electing a subset of candidates as a committee of representatives, and (2) the representatives voting sequentially on a fixed set of issues. This setting maps to most representative democracies across the world where representatives decide on a slate of issues between election rounds \cite{humanAttitudes}. In order to study pandering we analyze sequences of election cycles, and we refer to each of these election cycles as a round. 
The difference between RD and FRD is that in FRD the representatives vote using a weighted majority rule where the weights are determined by the voters, while in RD the representatives use (unweighted) simple majority rule on each issue.

\paragraph{Contributions}
We formalize and study a novel model of election attack, \emph{pandering}, where candidates report their positions strategically. We analyze two democratic voting systems, representative democracy (RD) and flexible representative democracy (FRD), in terms of their resilience to attack by pandering. We first show that the pandering is computationally hard for a single round, and provide an optimization program to solve this problem. We then model the problem of pandering over multiple rounds as a sequential decision making problem, formally a Markov Decision Problem (MDP). We then use techniques from reinforcement learning to solve this problem for pandering candidates and investigate how robust RD and FRD are to these attacks. We find that, generally, delegative voting systems such as FRD are more robust to these types of attacks.

\section{Related Work}


\subsection{Computational Social Choice}

Research in COMOSC focuses on computational aspects of collective decision making problems including voting and allocating resources among groups of agents who are self-interested \cite{brandt2016handbook}. Work in COMSOC has considered a variety of election attacks including manipulation, where a central agent is able to modify the votes of particular agents; bribery, where modification actions come at some cost and the attacker is bounded by a budget; and control, where one can remove or change the candidates of an election \cite{faliszewski2010ai}. While there are many results on both the complexity of and algorithms for these decision problems, they are typically studied in a single round. Most closely related to our work here is work on multi-issue \cite{binkele2014complexity} and shift bribery in committee elections \cite{bredereck2016complexity}, where an agent may pay to switch the preferences of individual voters. However, the algorithms for bribery problems are typically minimization problems under a budget constraint in a single round.

There has been some work in COMSOC involving strategic agents and multiple rounds including iterative voting \cite{meir2018strategic}, where agents repeatedly vote until a consensus is reached. The framework of dynamic social choice \cite{parkes2013dynamic} formulates the preferences of voters over candidates as an MDP and then investigates stationary policies in this setting as social choice functions, similar to e.g., page rank. However, neither of these settings involve strategic actions on the part of the candidates, only the voters. Most closely related to our work here is that of \citet{dutta2001strategic} who investigate strategic candidacy games, where agents can decide to participate or not in an election round as \emph{candidates}, which corresponds to a dynamic version of the control problem discussed above. However, in this setting candidates have fixed, known positions and are only deciding whether or not to stand for an election if they can win, whereas our candidates may misrepresent their positions.

We are focused on representative forms of democracy, where a small set of agents is selected from among a group of candidates. These problems are also studied within the COMSOC literature under the heading of committee elections or multi-winner elections \cite{chamberlin1983representative} and the questions around strategic attacks mentioned above have been studied for multi-winner elections \cite{procaccia2007multi}, but this work again does not investigate an entire democratic system, where later issues are to be decided, nor these questions over multiple rounds. 
%
Finally, Liquid Democracy \cite{blum2016liquid,brill2018interactive}, and variants including flexible/weighted representative democracy \cite{abramowitz2019flexible,weighted}, are popular areas of study in COMSOC as they provide a rich problem space to investigate as the underlying delegation graphs can be complex, and interconnected \cite{golz2021fluid}. We build directly on these systems and investigate novel election attacks within these systems. Finally, investigating the properties of voting systems over multiple rounds of decisions is becoming an area of interest within COMSOC, with ideas like perpetual voting \cite{lackner2020perpetual}, where properties like fairness should be ensured over election histories, have been recently studied.

\subsection{Sequential Decision Making, Reinforcement Learning and Security Games}

Sequential decision making and control problems are popular across AI. Many complex decisions must be made repeatedly, in the face of an uncertain and dynamic environment. The traditional tool to study these problems is the Markov Decision Process (MDP) \cite{sutton-barto-rl}. An MDP is a formal model where, over a number of time steps, $t \in T$, and an agent receives an observation of the current state $s \in \mathcal{S}$, where $\mathcal{S}$ defines the total state space of the system, and the agent must select an action $a \in A$ for each $s$. The environment evolves according to a transition function $T$ which gives probabilities of transitioning from one state to the next, given an action. At each state, the agent receives a real-valued reward signal, $R$, and the goal of the agent is to accrue as much (discounted) reward as possible while acting in this environment. The goal in an MDP is to find a policy $\pi: \mathcal{S} \rightarrow \mathcal{P}(A)$, i.e., a mapping of states to actions. Ideally, we want to solve an MDP by finding a policy $\pi^*$ that maximizes the expected (discounted) reward over a sequence of actions. In the MDP literature, classical tabular methods are used to find $\pi^*$ including value iteration (VI) and Q-learning. Such method finds an optimal policy by estimating the expected reward for taking an action $a$ in a given state $s$, i.e., the $Q$-value of pair $(s,a)$ \cite{sutton-barto-rl}. MDPs of this form are used across AI for sequential decision making tasks including recommending items to users \cite{ie2019reinforcement}, robot control \cite{abbeel2006application}, creating safe AI systems \cite{noothigattu-2019-ethicalvalues}, and modeling the dynamics between attackers and defenders in security games \cite{tambe2011security,lowe2017multi}.

Currently, deep reinforcement learning, which leverages deep learning for solving complex reinforcement learning problems with very large state and action spaces, achieves human level or above human-level performance on many tasks. Deep neural network enables reinforcement learning to approximate and parameterize the Q-table or other tabular values instead of computing them directly. Thus, deep reinforcement learning is capable of solving games with large state and action spaces. Classical deep reinforcement learning algorithms such as DQN \cite{mnih2013playing} conquer Atari games (which are video games with discrete action space such as Pong) and Go \cite{silver2016mastering}. Another algorithm PPO \cite{ppo} is capable of beating world champions in more complex video games such as DOTA2 \cite{berner2019dota}.

Sequential games have been commonly used to model strategic and learning behavior by agents in security settings. For example, an advanced and persistent attack typically starts with information collection to identify the vulnerability of a system and may act in a ``low-and-slow'' fashion to obtain long-term advantages~\cite{flipit,graph-apt}. On the other hand, an intelligent defender can profile the potential attacks and proactively update the system configuration to reverse information asymmetry~\cite{liammas2020,sengupta2020multi}. As the interaction between the attacker and the defender can generally be modeled as a Markov game with partial observations, reinforcement learning has been used to develop strong attacks and defenses in various settings. In particular, it has been used to corrupt the state signals received by a trained RL agent~\cite{robust-state-perturbation} and deceive a learning defender in repeated games~\cite{deception-security}. In a recent paper, \citet{RL-model-poisoning} show that RL-based attacks can obtain state-of-the-art performance in poisoning federated learning, where a set of malicious insiders craft adversarial model updates to reduce the global model accuracy. Given the difficulty of collecting sufficient samples in security-critical domains, an offline model-based approach is often adopted. To our knowledge, using RL to develop strategic attacks against voting systems has not been considered.   

\section{Electoral Pandering Model}\label{sec:model}


\subsection{Preferences Over Issues and Candidates}\label{sec:prefs}

Let $V$ be a set of $n$ voters and $C$ be a disjoint set of $m$ candidates, voters elect a subset of candidates $D \subset C$ where $|D| = k$ to serve as a committee. The committee of representatives will then vote on a sequence of $r$ binary issues.
We assume that every voter and candidate has a binary preference over every issue.
For voter $v \in V$, we denote their preference vector by $\bm{v} \in \{0,1\}^r$. Similarly, for candidate $c \in C$, their preference vector is denoted $\bm{c} \in \{0,1\}^r$.
The collective preference profiles are denoted by $\bm{V}$ and $\bm{C}$.

With $m$ candidates, there are $\binom{m}{k}$ possible ways to elect $k$ representatives.
However, it is infeasible for voters to express preferences over all possible committees of size $k$.
Therefore, representatives are elected via $k$-Approval with random tie-breaking. Each voter reports the subset of candidates of which they approve and the $k$ candidates who receive the greatest number of approvals get elected.
Following~\citet{abramowitz2019flexible}, voters submit approval preferences over candidates based on the fraction of issues on which they agree.
That is, $v$ approves of $c$ if $g(v, c) > \nicefrac{1}{2}$ where $g(v,c)$ is based on the Hamming distance between preference vectors. Let $d_H(\bm{x}, \bm{y}) = \sum_{i \leq r} |\bm{x}(i) - \bm{y}(i)|$ be the Hamming distance between two vectors of length $r$. For any two vectors $\bm{x}$ and $\bm{y}$ of length $r$, we refer to $g(\bm{x}, \bm{y}) = 1 - \frac{1}{r} d_H(\bm{x}, \bm{c})$ as their \emph{agreement} and $\frac{1}{r}d_H(\bm{x}, \bm{y})$ as their \emph{disagreement}. Intuitively, $g(v, c)$ is the fraction of issues the voter and candidate agree upon. Our measure of the quality of a voting system is the agreement (or disagreement) between the vector of outcomes it produces and the outcomes preferred by the voter majority.

\subsection{Pandering in Elections}

We introduce a novel model of election attack we call \emph{pandering}, wherein candidates are allowed to strategically misreport their private preferences in an attempt to get elected. 
%
We denote by $\hat{\bm{c}}$ the reported preferences of $c$, while their true preferences $\bm{c}$ remain private.
We assume that a subset of the candidates $S \subseteq C$ are strategic, i.e., candidate $c \in S$ may pander $(\bm{c} \neq \hat{\bm{c}})$. All other candidates $c \in C \backslash S$ are truthful.
Voter preferences over candidates are therefore based on the agreement between their preferences and the candidates' reported preferences: $\hat{g}(v, c) = 1 - \frac{1}{r} d_H(\bm{v}, \hat{\bm{c}})$.
Strategic candidates are assumed to know the full voter profile $\bm{V}$ but not the private or public preferences of other candidates at the time they report their public preferences. These strategic candidates pander in order to \emph{maximize} the number of approvals they receive to maximize their chances of being elected and hence affect the outcomes of the democratic system.
While candidates can be strategic about the preferences they report before the election, we assume they always vote according to their true preferences.
A more sophisticated candidate might also be strategic about when they vote according to their true preferences, but as we will show, computing one's pandering strategy when always voting according to one's true preferences is already NP-Hard.

\subsection{RD and FRD}
Following \cite{abramowitz2019flexible}, in classic Representative Democracy (RD) the candidates are elected by $k$-Approval with random tie-breaking, and each set of elected representatives votes on a sequence of $r$ binary issues using simple majority voting before the next election.
By contrast, in Flexible Representative Democracy (FRD) the representatives use weighted majority voting on every issue and these weights are determined on every issue by the voters. Each voter has 1 unit of weight to assign to the representatives and may distribute it among the representatives however they wish. The weight of a representative on an issue is then the sum of weights assigned to them.
That is, each voter $v$ assigns each representative $c$ a weight $0 \leq w^t(v,c) \leq 1$ on each issue $t$ such that $\sum_{c \in D}w^t(v,c) = 1$ for all $t$ and the weight of a representative is $w^t_c = \sum_{v \in V} w^t(v,c)$. If $\bm{c}(t) \in \{0,1\}$ is the preference of $c$ on issue $t$, then weighted majority voting leads the outcome to be 1 if $\sum_{c \in D}w^t_c \bm{c}(t) > \nicefrac{n}{2}$, 0 if $\sum_{c \in D}w^t_c \bm{c}(t) < \nicefrac{n}{2}$, and breaks ties randomly otherwise.
Section~\ref{sec:multiround_model} will detail how we model the way voters assign these weights in our pandering model.

\section{Pandering in a Single Round}

We show that even in a single round it is NP-Hard for a strategic candidate $c \in S$ to compute the profile $\hat{\bm{c}}$ that maximizes the number of approvals they receive when $c$ has full information about the voter preferences $\bm{V}$. We care maximizing approvals as we do not assume they have access to the reported preferences of other candidates at the time they report their own preferences.

\begin{problem}[Maximum Approval Pandering (MAP)] \label{prob:map}
    Given a profile of $n$ voters over $r$ issues $\bm{V} \in \{0,1\}^{r \times n}$, compute $\hat{\bm{c}} = \argmax\limits_{\bm{c} \in \{0,1\}^r} |\{\bm{v} \in \bm{V}: d_H(\bm{v}, \bm{c}) < \frac{r}{2}\}|$.
\end{problem}
\vspace{-1em}

Our proof below that Maximum Approval Pandering is NP-Hard follows a proof by Neal Young~\cite{mapStackExchangeNealYoung}, with slight modification and simplification. The proof uses a Karp reduction via the known NP-Complete problem of Max 2-SAT~\cite{garey1979computers}. In Max 2-SAT, one is given a Boolean formula in conjunctive normal form where each clause contains at most two literals and the task is to find an assignment to the variables such that a maximum number of clauses is satisfied.

\begin{theorem}\label{thm:MAP_hard}
    Maximum Approval Pandering is NP-Hard
\end{theorem}

\begin{proof}
    Suppose we have a Boolean formula in conjunctive normal form with $n$ variables and $m$ clauses for which each clause has exactly two literals. Assume without loss of generality that $n = 2^k$ is some power of 2.
    We will construct a collection of binary vectors $\bm{V}$ to serve as input to an instance of MAP.
    We start by adding $m+1$ copies of the vector $(0)^{2n}$ and $m+1$ copies of $(1)^{2n}$ to the collection $\bm{V}$.
    Consider the elements of each vector $\bm{v} \in \bm{V}$ to be in pairs so that $\bm{v}$ is of the form $\{00, 01, 10, 11\}^n$.
    Now for all $j \in \{2^i : 0 \leq i < k\}$, add $m+1$ copies of the string $(0^j 1^j)^{n/j}$ and $m+1$ copies of its complement $(1^j 0^j)^{n/j}$.
    Now $\bm{V}$ contains $2k(m+1)$ vectors, each of length $2n$.
    Notice that for a vector $\bm{c}$ to be within a distance $d_H(\bm{v}, \bm{c}) \leq n$ of all vectors $\bm{v} \in \bm{V}$, it must be of the form $\{01, 10\}^n$. Any other vector $\bm{c}$ will have $d_H(\bm{c}, \bm{v}) > n$ for at least $m+1$ of the vectors in $\bm{V}$.
    Now we add $m$ additional vectors to $\bm{V}$ based on the clauses of our Boolean formula.
    For each clause, let $x_i$ and $x_j$ be the two variables that appear in the clause, and construct the vector $\bm{v}$ such that all elements are zero, except that the $i^{th}$ (resp. $j^{th}$) pair is $01$ if $x_i$ (resp. $x_j)$ appears positively in the clause and $10$ if it appears negatively.
    Thus, $\bm{V}$ now contains $m$ additional binary vectors each of length $2n$, and each contains exactly 2 ones and $2n-2$ zeroes.
    Any vector $\bm{c}$ that maximizes the number of vectors $\bm{v} \in \bm{V}$ for which $d_H(\bm{c}, \bm{v}) \leq n$ must still be of the form $\{01, 10\}^n$, because a different vector could reduce it's distance to the $m$ new vectors based on clauses only at the expense of being too great a distance from at least $m+1$ of the other vectors.
    As $v$ only approves of $c$ if they agree on strictly more than half the issues, not greater than or equal to half the issues, append a 1 to all vectors in $\bm{V}$. Now any solution to the MAP instance will be of the form $(\{01, 10\}^n)(1)$ and its first $n$ pairs of values $01$ and $10$ can be read as giving the truth values of the variables in the original Boolean formula.
\end{proof}

The complexity of MAP may be surprising, as one might expect that a candidate taking the position of the voter majority on each issue would be optimal. However, Anscombe's Paradox shows in dramatic fashion that this is not the case, as for certain voter profiles the majority of voters can be in the minority on the majority of issues~\cite{anscombe1976frustration}.
We will use this greedy pandering strategy of reporting the voter majority preference on every issue as a baseline for comparison in Section \ref{sec:experiments}. We will discuss pandering optimally in a single round in more detail in Section \ref{sec:singleround}.

\section{Pandering in Multiple Rounds}\label{sec:multiround_model}
If we only considered a single round, strategic candidates would pander on as many issues as necessary to maximize the number of approvals they receive without consequence since voters would not discover the strategic actions and then distrust that candidate.
Hence, we extend our setting into a \emph{multi-round model} where strategic candidates face consequences for past pandering, since these actions hurt their \emph{credibility} in the eyes of the voters.
We now focus on sequences of election cycles, or \emph{rounds}, in which committees of representatives are elected at regular intervals. We assume that number of issues $r$ is the same for all rounds.

A time step $t$, there have been $t-1$ issues already decided, and the next issue to be voted on by the representatives is issue $t$. 
Agent preferences over singular issues are indexed as $\bm{v}(t)$ and $\bm{c}(t)$ respectively.
Some time steps correspond to the beginning of a new round in which an election must take place before issue $t$ is decided. We will use $q_t$ to denote the round containing time step $t$. 
We use the superscript $q$ to denote variables defined for round $q$ including the preference profiles $\bm{V}^q$ and $\bm{C}^q$ over only the issues of that round, individual preferences $\bm{v}^q$ and $\bm{c}^q$ which are binary strings of length $r$, the set of elected representatives $D^q \subset C$, and the fraction of issues agreed upon by a voter and candidate in that round $\hat{g}^q(v,c)$.
While strategic candidates may misreport their preferences to get elected, we assume that they always vote according to their true preferences once they have been elected to the representative body. All pandering by representatives is revealed, but not the pandering of candidates who do not get elected.

At every time step, each candidate has a credibility $0 \leq h_c(t) \leq 1$, where initially $h_c(1) = 1$ for all candidates implying a presumption of total honesty at the start. 
The credibility of a candidate can never become greater than $1$. We denote the credibility of a candidate at the time of an election by $h_c^q$, so if $t$ is the first time step of a new round then $h_c^{q_t} = h_c(t)$.
Now, in each election, voter $v$ approves of candidate $c$ in round $q$ if and only if $\hat{g}^q(v,c) h_c^q > \nicefrac{1}{2}$. That is, voters' approvals depend on both their agreement with a candidate and how credible the candidate is. Even if a voter agrees with a candidate on every issue, if the candidate is not sufficiently credible, the voter will not approve of them.

In RD, the credibility of candidates only matters at the beginning of each round, when the voters express their preferences over the candidates, as once a candidate is elected, all issues are decided independently by the representatives until the next round. However, for FRD, the credibility of representatives affects how the voters weight them on each issue, and so the way their credibility updates at each time step matters.
In FRD, the representatives decide each issue by weighted majority vote, where the issue-specific weights are determined by the voters.
The weights assigned by a voter to the elected representatives must sum to 1, so all voters contribute an equal total weight.
We assume that voters assign weight only to representatives who agree with them on each issue, and assign it in proportion to the representatives' credibility.
The weight assigned by $v$ to representative $c \in D$ on issue $t$ is $$w^t(v,c) = \frac{(1-|\bm{v}(t) - \bm{c}(t)|) h_c(t)}{\sum\limits_{c \in C} (1-|\bm{v}(t) - \bm{c}(t)|) h_c(t)}$$
The weight of a representative on issue $t$ is the sum of weights assigned to them: $w^t_c = \sum_{v \in V} w^t(v,c)$.

We model three competing forces influencing the credibility of candidates over time: changes in credibility for pandering, changes for being truthful, and for un-elected candidates, changes due to the fading memory of past pandering.
If a candidate gets elected, their credibility is updated in each time step after they vote. If a candidate is not elected, their credibility is updated before the next round.
If $c$ is elected in round $q_t$ and panders on issue $t$, then $h_c(t+1) = \beta_1 h_c(t)$ where $0 \leq \beta_1 \leq 1$ reflects how sensitive the voters are to pandering revelations. If $c$ is elected and does not pander on issue $t$, then $h_c(t+1) = \max\{(1+\beta_2) h_c(t), 1\}$, where $\beta_2 \geq 0$ reflects how much credibility a candidate earns by being truthful on an issue.
Lastly, if a candidate is not elected, it is never revealed to what degree they pandered in that round and so their credibility is updated at the end of the round by $h_c^{q+1} = \max\{1, h_c^q (1+\beta_3)\}$ where $\beta_3 \geq 0$ reflects the fading memory of their past pandering.

Given that it is NP-Hard for to solve MAP in a single round it is at least as hard for a candidate to be optimally strategic over multiple rounds when agent preferences in future rounds are not known in advance and effects on the candidate's credibility over time must be taken into account. Hence, for sequential decision making problem, we turn to reinforcement learning to study how effective a candidate can be in pandering over many rounds.


\subsection{Voting Systems as MDPs}
In our analysis we consider two types of strategic candidates: selfish and malicious.
A \emph{selfish candidate} seeks to maximize their influence over the outcomes to push them in favor of their preference, so their utility is based on the number of issues they cause to agree with their personal preference as a representative.
On the other hand, a \emph{malicious candidate} prefers the opposite outcome to the voter majority on every issue (they just want to watch the world burn). Their utility is based on the number of issues whose outcome disagrees with the voter majority, i.e., the total disagreement. 
In our experiments we also investigate the robustness of these systems when there are groups of strategic candidates. Malicious candidates coordinate with one another, while selfish candidates do not.
%
%
%
%
In the next section we formally define the decision problem of the candidates as a finite horizon MDP, where the horizon is 100 rounds, each containing 9 issues. We use a discount factor $\gamma = 1$, i.e., no discounting of future rewards in all our analysis.

\begin{figure*}[ht]
    \centering
   \begin{tabular}{cccc}
    \adjustbox{valign=m}{{%
    \subfigure[]{
          \centering
          \includegraphics[width=.21\textwidth]{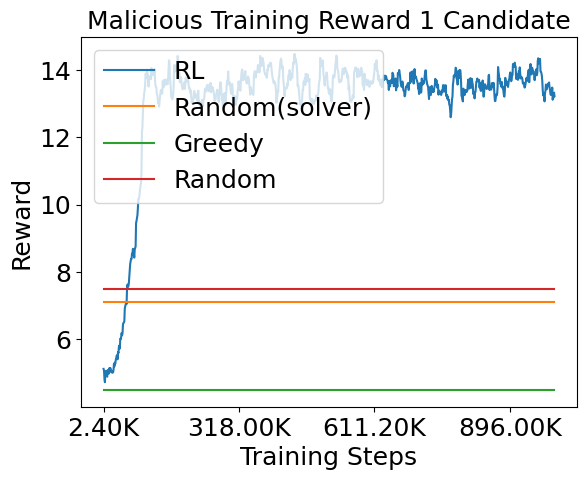}
          }
          }}
         &
    \adjustbox{valign=m}{{%
    \subfigure[]{
          \centering
          \includegraphics[width=.21\textwidth]{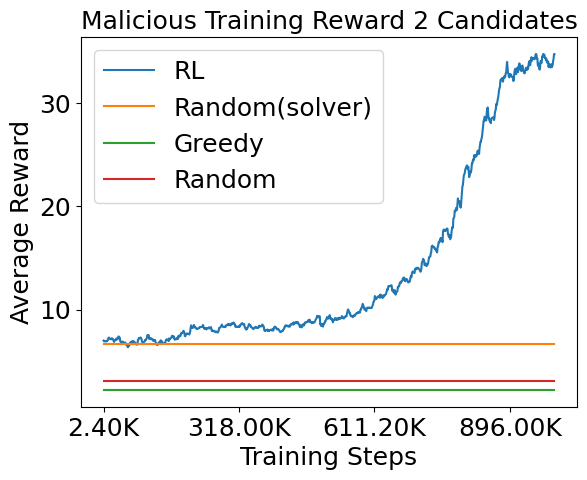}
          }
          }}
         &
    \adjustbox{valign=m}{{%
    \subfigure[]{
          \centering
          \includegraphics[width=.21\textwidth]{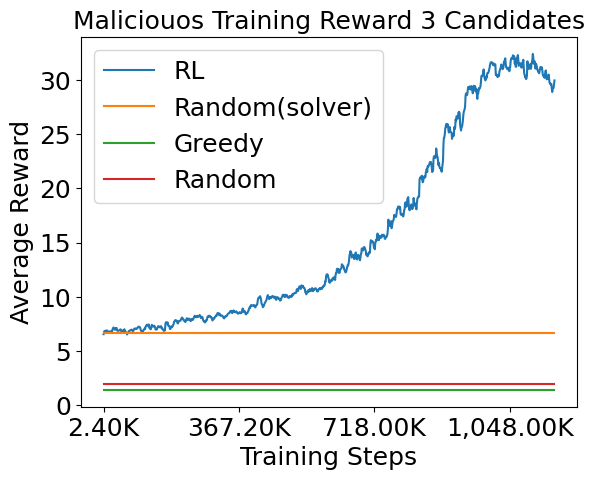}
          }
          }}
         &
    \adjustbox{valign=m}{{%
    \subfigure[]{
          \centering
          \includegraphics[width=.21\textwidth]{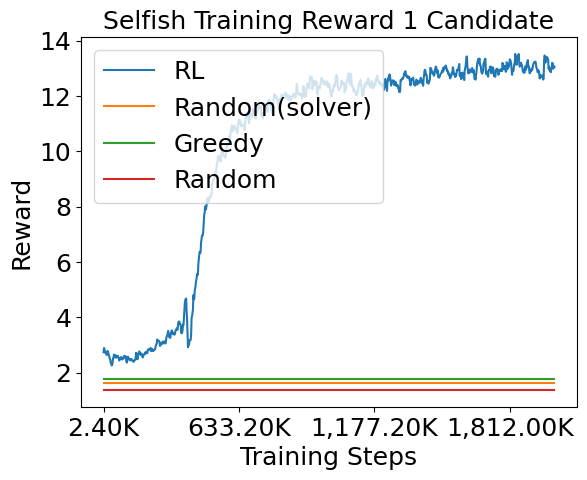}
          }
          }}\\
    \end{tabular}
    \caption{\small{Training curve of strategic candidates in RD voting systems compared with baselines, $\beta_1 = 0.95$.
    }}\label{fig:convergence}
    \vspace{-1em}
  \end{figure*}

\subsection{MDP for Selfish Candidates}
Since selfish candidates only care about their own benefit, each of them will take actions independently from each other. For each selfish candidate $c \in S$ we define: the \emph{state space} as $\mathcal{S}^q = (\bm{V}^q, \bm{c}^q, h_{c}^q, q)$; the \emph{action space} as $A^q = \{0,1\}^r$; and the \emph{transition function} as $T: \mathcal{S}^q \times A^q \rightarrow \mathcal{P}(\mathcal{S}^{q+1})$. We assume that all voters' and candidates' preferences are i.i.d. random variables from a fixed stationary distribution across all rounds. Therefore, the probability of any voter and candidate profile is independent of the state and history. Similarly, the credibility of candidates at the beginning of the round $h^{q+1}_c$ depends only on $h^{q}_c$ and $A^{q}$, so the Markov property is satisfied. 

The goal of an individual selfish candidate is to find a policy $\pi: \mathcal{S}^q \rightarrow A^q$ in order to maximize the cumulative reward over a finite time horizon (100 rounds). We set the \emph{reward} as $R^q = {f(a^q,\bm{c}^q, D^q(c),\bm{o}^q)}$, where $D^q(c)$ is a binary indicator variable representing whether the strategic candidate $c$ is elected in round $q$ only through pandering, which means the strategic candidate $c$ will not get elected by being honest in round $q$, and $\bm{o}^q$ is the binary vector of outcomes in round $q$. Informally, if a selfish is elected due to pandering when they otherwise would not have been, their reward is equal to the number of issues on which they agree with the outcome in that round. Otherwise, if they do not get elected or would be elected by being honest, the reward is zero.

\subsection{MDP for Malicious Candidates}
Since malicious candidates share the same objective, they cooperate with each other in order to damage the system. Thus, we model all malicious candidates as sharing the same state space, action space, and reward. Formally, the \emph{state space} is $\mathcal{S}^q = (\bm{V}^q, \bm{c}^q, \{h_{c'}^q\}_{c' \in S}, q)$; the \emph{action space} is $A^q = \{0,1\}^{r |S|}$; and \emph{transition function} $T: O^q \times A^q \rightarrow \mathcal{P}(\mathcal{S}^{q+1})$ is the state transition function, which is the same as the selfish candidate MDP described above.

The goal of the malicious candidates is to find a joint policy $\pi: \mathcal{S}^q \rightarrow A^q$ in order to maximize the joint cumulative rewards over the time horizon. We set the \emph{reward function} of the malicious candidates to be $R^q = {f(\bm{o}^q, \tilde{\bm{o}}^q)}$ where $\tilde{\bm{o}}^q$ is the vector of outcomes that would have resulted if no strategic candidates pandered in round $q$. Informally, malicious candidates only care about increasing disagreement. If a malicious candidate is elected due to pandering when they otherwise would not have been elected by being honest, they receive a reward equal to the number of issues whose outcomes disagree with the voter majority minus the number of issues that would have disagreed with the voter majority had they been truthful. Thus, if a malicious agent is not elected, is elected by being truthful, or the outcomes all agree with the voter majority, the agent receives a reward of zero.

\section{Experiments}\label{sec:experiments}


\subsection{Pandering in a Single Round}\label{sec:singleround}

\begin{figure}[h]
    \centering
    \includegraphics[width=.22\textwidth]{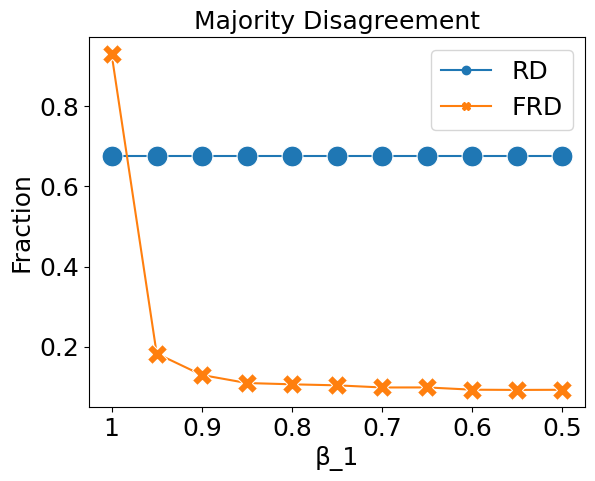}
    \caption{Fraction of outcomes that disagree with the voter majority for a single round with 900 issues vs. $\beta_1$ with a single malicious candidate pandering greedily.}
    \label{fig:oneround}
\end{figure}

In Maximum Approval Pandering (MAP), if we were only interested in a single round, then candidates would pander greedily, on as many issues as possible in order to get elected. In Figure \ref{fig:oneround} we plot the fraction of disagreement, i.e., how often the outcome of the election system agrees with the voter majority for a single round with 900 issues. The malicious candidate panders by reporting their preferences $\bm{c}$ as equal to the voter majority on every issue (greedy), whereas their private preferences $\hat{\bm{c}}$ is actually the voter minority on every issue. In this simulation, and in all experiments in our paper, the preferences of all voters and truthful candidates are uniformly random on every issue, i.e., $p = 0.5$. 

Even though the malicious candidate panders on all issues, Figure \ref{fig:oneround} already illustrates some interesting differences between RD and FRD. Notably, if voters ignore the pandering of candidates ($\beta_1 = 1$) for a single round, then one is better off not allowing a weighting of the representatives and sticking with RD instead. This is because strategic candidates will be given higher weight, as their reported preferences were better able to match those of the voters. However, once voters pay attention to the pandering of candidates and begin to punish them for it even slightly, FRD becomes far superior to RD in following the voter majority.

\begin{figure*}[ht]
    \centering
   \begin{tabular}{cccc}
    \adjustbox{valign=m}{{%
    \subfigure[]{
          \centering
          \includegraphics[width=.21\textwidth]{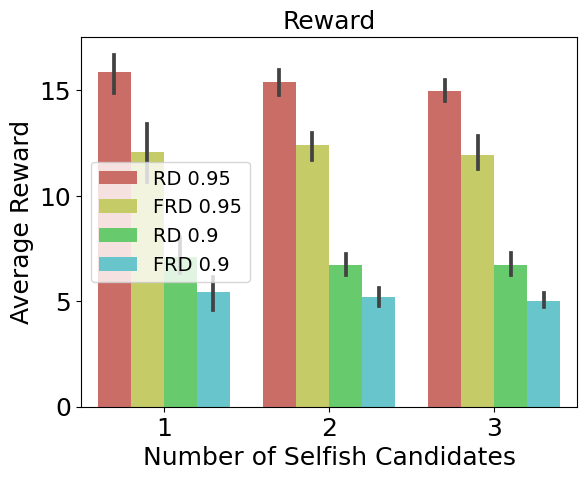}
          }
          }}
         &
    \adjustbox{valign=m}{{%
    \subfigure[]{
          \centering
          \includegraphics[width=.21\textwidth]{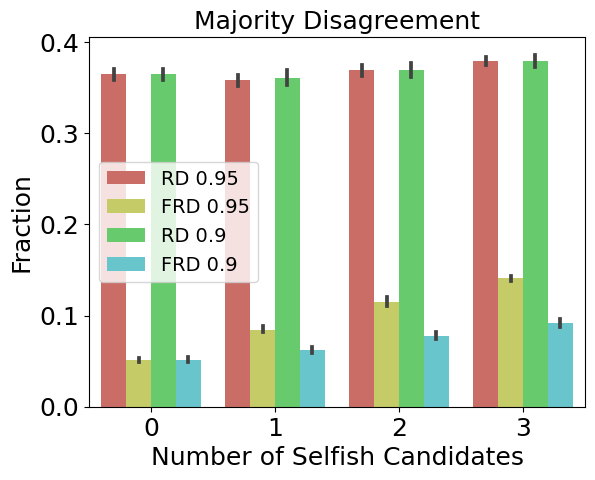}
          }}}
    \adjustbox{valign=m}{{%
    \subfigure[]{
          \centering
          \includegraphics[width=.21\textwidth]{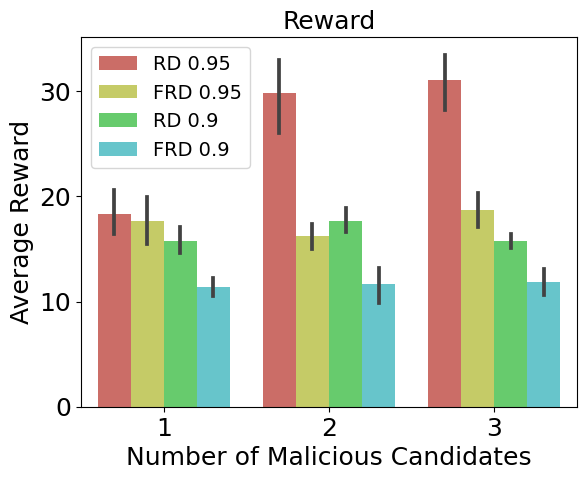}
          }
          }}
         &
    \adjustbox{valign=m}{{%
    \subfigure[]{
          \centering
          \includegraphics[width=.21\textwidth]{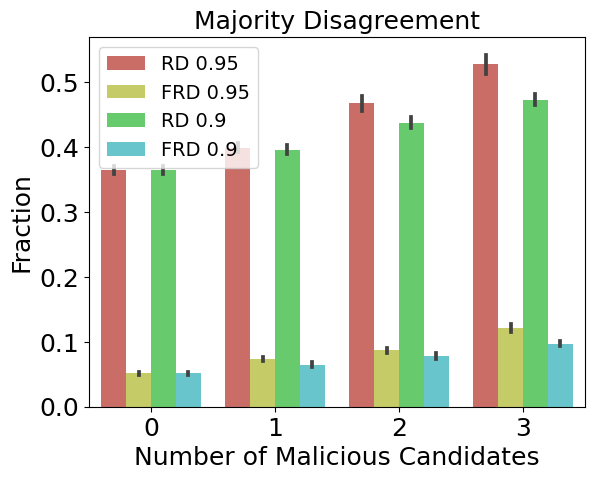}
          }}}
    \end{tabular}
    \caption{\small 
    Effects of pandering by up to $|S| = 3$ selfish candidates out of $|C| = 10$ in RD and FRD where $k = 5$ get elected. Figure (a) shows the average reward of each selfish candidate for $\beta_1 \in \{0.9, 0.95\}$. Figure (b) shows what fraction of the 900 total issues (across 100 rounds) are decided against the voter majority for $\beta_1 \in \{0.9, 0.95\}$. Malicious candidates for the same settings are shown in (c) and (d).
    }\label{fig:combined_results}
    \vspace{-1em}
  \end{figure*}

\subsection{Multiple Round Setup}

\paragraph{Preferences} In the rest of our experiments, the preferences of all voters, truthful candidates, and selfish candidates are drawn uniformly at random for every issue, i.e., $p = 0.5$. We set the preferences of malicious candidates to be the voter minority on every issue as they are seeking to create the most disagreement possible. We refer the reader to the appendix to view experiments with different preference distributions.

\paragraph{Reinforcement Learning and Environment Details} In each round we have $n = 50$ voters, $|C| = 10$ candidates, of which $|D| = 5$ will be elected. We vary the number of strategic candidates $|S| \in \{0, 1, 2, 3\}$. In each round the representatives will vote on a sequence of $r=9$ binary issues, and there will be 100 rounds for a total of 900 issues (time horizon). For the credibility parameters we examine $\beta_1 \in \{0.9, 0.95\}$ based on our findings Figure\ref{fig:oneround}, and fix $\beta_2 = 0.003$ and $\beta_3 = 0.01$ for simplicity. We use the DQN algorithm implemented with stable-baselines3~\cite{stable-baselines3} to all our agents. Either all of the candidates in $S \subset C$ are selfish or they are all malicious. We train policies for one, two and three malicious candidates and policy for one selfish candidate. We ran our experiments on a server with Intel i9-12900KF and Nvidia RTX 3090. It took about 60 hours to train 1 million rounds for three malicious candidates.

\paragraph{State Space Compression} The dimension of the state space in the MDP for selfish candidates is $(n+1)\cdot r+2$ and for malicious candidates it is $(n+1)\cdot r+|S|+1$. These dimensions are too large to efficiently train our candidates. In order to compress the large state space, we compress the full profile $\bm{V}^q$ down to the vector $\bm{v}^{*q}$ where $\bm{v}^{*q}(t) = \frac{1}{r}\sum_{v \in V}\bm{v}^q(t)$. This compression decreases the dimension of voter preferences from $n \cdot r$ down to $r$ for both selfish and malicious candidates but at the cost of not knowing the preferences of any specific voter on any issue.

\paragraph{Action Space Compression} Even if candidates could solve the MAP problem in every round, it is not necessarily optimal to pander on as many issues as necessary to get elected, as some rounds may require more or less pandering. Candidates must be strategic about how many issues they are willing to lie about in a given round. Hence, we give our candidates the ability to solve the following more general version of the MAP problem in every round, so their only strategic choice is in selecting the maximum number of issues they are willing to pander on in each round.
    \begin{problem}[Constrained Maximum Approval Pandering (CMAP)]
        \label{mip}
        For any profile of voter preferences over $r$ binary issues $\bm{V} \in \{0,1\}^{r \times n}$, private preferences $\bm{c}$ of a candidate, and integer $0 \leq a \leq r$, compute a preference to report that maximizes approvals subject to the constraint that it panders on at most $a$ issues:
        $$\hat{\bm{c}} = \argmax\limits_{\bm{c}': d_H(\bm{c}, \bm{c}') \leq a} |\{\bm{v} \in \bm{V}: d_H(\bm{v}, \bm{c}') < \frac{r}{2}\}|$$
    \end{problem}

\noindent
We created a mathematical program to solve CMAP with Mathmatica~\cite{Mathematica} for all our experiments on pandering. With the CMAP problem, the action space becomes $A^q = \{0, 1, 2, \ldots, r\}$ for selfish candidates and $A^q = \{0, 1, 2, \ldots, r\}^{|S|}$ for malicious candidates.

\paragraph{Reward Design} 
Selfish candidates only receive reward if they get elected by pandering and not if they are being honest. This reward function encourages selfish candidate to find the best pandering policy that will let them get elected the most, which corresponds to real life selfish politicians who want to maximize their own fame. In the malicious setting, all malicious candidates share the same reward in each round which captures how far the malicious candidates, who want to devastate the voting system, can get the outcomes to deviate the majority will of the voters.
%
Selfish: For each $c \in S$: $R^q_c = f(a^q, c^q, D^q(c),\bm{o}^q) 
        = D^q(c) \cdot (1 - \frac{1}{r}d_H(\bm{o}^q, \bm{c}))$. 
Malicious:$R^q = {f(\bm{o}^q, \tilde{\bm{o}}^q)} = \frac{1}{r} d_H(\bm{o}^q, \tilde{\bm{o}}^q)$.
    

\paragraph{Testing Details} We run each of our experiments under 10 random seeds and report average and error bars in our graphs. Testing environments uses the same parameters as training environment. For the setting with multiple selfish candidates, each of the selfish candidate uses the same policy. This means that, from the view of each selfish candidate  all other selfish candidates are treated as benign candidates. This gives us an approximation of the optimal policy in this setting. We do this as solving the full Markov game, or multi-agent MDP induced by multiple, self-interested selfish candidates where each takes the others actions into consideration, is computationally infeasible.

\subsection{Multiple Round Results}

\subsubsection{Convergence and Baseline Comparison}

Figure \ref{fig:convergence} shows the training curve for varying numbers of malicious candidates and a single selfish candidate. Along with the training curves to show convergence, we plot three naive baselines: random, random(solver) and greedy. A random pandering candidate randomly chooses $\hat{\bm{c}}$ in each round, the random(solver) pandering candidate will randomly choose the number of issues to lie each round and feed the number into the CMAP solver to generate $\hat{\bm{c}}$, while an candidate that is greedily pandering always chooses the voter majority as his/her public preference.
Figure~\ref{fig:convergence} (a) shows that our RL candidate is able to quickly learn how to pander, and outperforms all of the baselines. Looking at (b) and (c) we see that as we add more malicious candidates, the convergence takes longer, but the malicious candidates are able to outperform the baselines by a greater margin as the candidates are able to learn a cooperative policy and achieve a higher reward. In fact, even at the start of training, the RL candidates are able to outperform the baselines, i.e., with very little training.


\subsubsection{Experiments with Selfish Candidates}
Figure~\ref{fig:combined_results} details the results of our experimental results with selfish candidates. Recall that selfish candidates pander to increase their influence over election outcomes and steer the voting outcomes to match their private preferences. Figure~\ref{fig:combined_results}(a) shows that the average reward received by selfish candidates is lower under FRD than it is under RD in every scenario, indicating that FRD is more resilient to pandering by selfish candidates. However, the difference between $\beta_1 = 0.9$ and $\beta_1 = 0.95$ dwarfs the difference between RD and FRD, meaning that sensitivity to pandering has a much greater effect on the reward of selfish candidates than allowing the weighting of representatives. 

Figure~\ref{fig:combined_results}(b) shows that FRD is significantly better than RD at leading to voting outcomes that have lower disagreement, i.e., represent the voter majority, no matter how many selfish candidates are present. Here again we see that the difference between RD and FRD is much larger than the difference between $\beta_1 = 0.9$ and $\beta_1 = 0.95$. To highlight the drastic difference, in RD with no selfish candidates at all over 30\% of issues decided by the representatives go against the voter majority, while in FRD with 3 selfish candidates and the weaker value of $\beta_1 = 0.95$, the fraction of issues decided against the voter majority is below 15\%.

  
\subsubsection{Experiments with Malicious Candidates}

Figure \ref{fig:combined_results} shows the results on average candidate reward and majority disagreement for settings with varying a varying number of malicious agents. 
Figure~\ref{fig:combined_results}(c) shows that FRD yields a lower average reward for malicious candidates than RD for any number of malicious candidates, but the difference between $\beta_1 = 0.9$ and $\beta_1 = 0.95$ is far less striking than for selfish candidates as seen in Figure \ref{fig:combined_results}(a). Thus, both sensitivity to pandering and the weighting of representatives are important in the presence of malicious candidates. 

Note that the reward functions are different for the two candidates types, so these scales are not directly comparable, only the relative effect sizes of the different parameters are. Figure~\ref{fig:combined_results}(d) shows a similarly drastic difference between FRD and RD that dwarfs the difference between the two $\beta_1$ values. The fraction of issues that disagree with the voter majority is higher for every number of strategic candidates when the candidates are malicious versus when they are selfish under RD, but for FRD there is little difference.

\section{Discussion and Conclusions}

As we have seen in Section \ref{sec:experiments}, FRD is able to account for the presence of both malicious and selfish candidates better than RD, resulting in more issues being decided with the majority of voters. Comparing the results in Figure \ref{fig:oneround} with those in Figure \ref{fig:combined_results}, we see that holding regular elections is, in fact, important in upholding the "will of the people". In Figure \ref{fig:oneround}, even for RD, the fraction of majority disagreement is around 0.68, while in Figure \ref{fig:combined_results} it is 0.5. So, even when we have multiple, optimal malicious candidates, they are unable to cause as much damage as a single candidate in a single round. This observation holds for both RD and, even more so for FRD.
We observe that FRD has lower fraction of disagreement and a lower attacker reward across all tested scenarios. Thus, we can draw the conclusion that FRD is more resilient than RD in the face of pandering. The average reward is almost the same for all scenarios expect malicious candidates under $\beta_1 = 0.95$, indicating that damage from strategic candidates is almost linear in $|S|$, except malicious when $\beta_1 = 0.95$, where we see that a high tolerance for pandering leads to more coordination opportunities for malicious candidates.




\section{Acknowledgements}
Nicholas Mattei was supported by NSF Awards IIS-RI-2007955, IIS-III-2107505, and IIS-RI-2134857, as well as an IBM Faculty Award and a Google Research Scholar Award. Ben Abramowitz was supported by the NSF under Grant \#2127309 to the Computing Research Association for the CIFellows Project. Xiaolin Sun and Zizhan Zheng were supported by NSF awards CNS-1816495 and CNS-2146548, and Tulane University Jurist Center for Artificial Intelligence. The authors thank Neal Young for his original proof of Theorem~\ref{thm:MAP_hard}~\cite{mapStackExchangeNealYoung} and to the anonymous user on StackExchange for their help in debugging~\cite{StackExchangeDebug}.


\bibliographystyle{plainnat}
\bibliography{refs}

\end{document}